\documentclass[12pt,showpacs,preprintnumbers,superscriptaddress,amsmath,amssymb,nofootinbib]{revtex4}
\usepackage{amsmath, graphicx}
\usepackage{dcolumn}
\usepackage{bm}
\usepackage{epstopdf}
\begin{document}
\newcommand{\hs}{\hspace*{0.5cm}}
\newcommand{\vs}{\vspace*{0.5cm}}
\newcommand{\be}{\begin{equation}}
\newcommand{\ee}{\end{equation}}
\newcommand{\bea}{\begin{eqnarray}}
\newcommand{\eea}{\end{eqnarray}}
\newcommand{\ben}{\begin{enumerate}}
\newcommand{\een}{\end{enumerate}}
\newcommand{\bde}{\begin{widetext}}
\newcommand{\ede}{\end{widetext}}
\newcommand{\nn}{\nonumber}
\newcommand{\crn}{\nonumber \\}
\newcommand{\Tr}{\mathrm{Tr}}
\newcommand{\non}{\nonumber}
\newcommand{\noi}{\noindent}
\newcommand{\al}{\alpha}
\newcommand{\la}{\lambda}
\newcommand{\bet}{\beta}
\newcommand{\ga}{\gamma}
\newcommand{\va}{\varphi}
\newcommand{\om}{\omega}
\newcommand{\pa}{\partial}
\newcommand{\+}{\dagger}
\newcommand{\fr}{\frac}
\newcommand{\sq}{\sqrt}
\newcommand{\bc}{\begin{center}}
\newcommand{\ec}{\end{center}}
\newcommand{\Ga}{\Gamma}
\newcommand{\de}{\delta}
\newcommand{\De}{\Delta}
\newcommand{\ep}{\epsilon}
\newcommand{\varep}{\varepsilon}
\newcommand{\ka}{\kappa}
\newcommand{\La}{\Lambda}
\newcommand{\si}{\sigma}
\newcommand{\Si}{\Sigma}
\newcommand{\ta}{\tau}
\newcommand{\up}{\upsilon}
\newcommand{\Up}{\Upsilon}
\newcommand{\ze}{\zeta}
\newcommand{\ps}{\psi}
\newcommand{\Ps}{\Psi}
\newcommand{\ph}{\phi}
\newcommand{\vph}{\varphi}
\newcommand{\Ph}{\Phi}
\newcommand{\Om}{\Omega}

\title{Lepton mass and mixing in a simple extension \\
 of the Standard Model based on $T_7$ flavor symmetry}
\author{V. V. Vien}
\email{wvienk16@gmail.com} \affiliation{Institute of Research and Development, Duy Tan University,\\ 182 Nguyen Van Linh, Da Nang City, Vietnam}
\affiliation{Department of Physics, Tay
Nguyen University, \\
567 Le Duan, Buon Ma Thuot, DakLak, Vietnam}

\author{H. N. Long}
\email{hnlong@iop.vast.ac.vn}
 \affiliation{Theoretical Particle Physics and Cosmology Research Group and Faculty of Applied Sciences, Ton Duc Thang University, Ho Chi Minh City, Vietnam}
 \affiliation{Institute of Physics,
VAST, 10 Dao Tan, Ba Dinh, Hanoi, Vietnam.}
\date{\today}

\begin{abstract}

 A simple Standard Model Extension based on $T_7$ flavor symmetry which  accommodates lepton mass and  mixing with non-zero $\theta_{13}$ and CP violation phase
is proposed. At the tree- level,  the realistic lepton mass and mixing pattern is derived through  the spontaneous symmetry breaking by just one vacuum expectation value ($v$) which is the same as in the Standard Model. Neutrinos get small masses from one $SU(2)_L$ doublet and two $SU(2)_L$ singlets  in which one being in $\underline{1}$ and the two others
in $\underline{3}$ and $\underline{3}^*$ under $T_7$ , respectively. The model also gives a remarkable prediction of Dirac CP violation $\delta_{CP}=172.598^\circ$
in both normal and inverted hierarchies which is still missing in the neutrino mixing matrix.

\end{abstract}

\maketitle

\section{\label{intro} Introduction}

The discovery of neutrino mass is  a great breakthrough for particle physics, and up to now,  this is the unique evidence
of New Physics. Neutrinos have tiny masses and this is probably related to the existence of  a new mass scale in physics.  Recently it has been shown that neutrinos can also play an important role in providing answer for the Baryon Asymmetry of Universe (BAU).

Theoretically, there exist various models describing the smallness of neutrino mass and large
 $\theta_{13}$ mixing\footnote{The references for these models  are mentioned in Ref. \cite{VienS4SM}}. Among the possible extensions of the Standard Modem (SM), probably the simplest one is the neutrino minimal SM  which has been studied in  Refs.  \cite {nuMSM1,nuMSM2,nuMSM3,nuMSM4,nuMSM5}. However, these extensions do not provide a natural explanation for large mass splitting between neutrinos and the lepton mixing was not explicitly explained \cite{TA} .

There are five well-known patterns of lepton mixing \cite{Xing2012}, however, the Tri-bimaximal one proposed by
Harrison-Perkins-Scott (HPS) \cite{hps1, hps2, hps3, hps4}
\be
U_{\mathrm{HPS}}=\left(
\begin{array}{ccc}
\frac{2}{\sqrt{6}}       &\frac{1}{\sqrt{3}}  &0\\
-\frac{1}{\sqrt{6}}      &\frac{1}{\sqrt{3}}  &\frac{1}{\sqrt{2}}\\
-\frac{1}{\sqrt{6}}      &\frac{1}{\sqrt{3}}  &-\frac{1}{\sqrt{2}}
\end{array}\right),\label{Uhps}
\ee
 seems to be the most popular and can be considered as a leading order approximation for the recent
neutrino experimental data. Up to now, the absolute values of the entries of the lepton mixing matrix
$U_{PMNS}$ have not yet been determined exactly, however, their scales are given in Ref. \cite{Gonzalez2014}
\bea
\left|U_{\mathrm{PMNS}}\right|=\left(
\begin{array}{ccc}
 0.801 \to 0.845 &\qquad
    0.514 \to 0.580 &\qquad
    0.137 \to 0.158
    \\
    0.225 \to 0.517 &\qquad
    0.441 \to 0.699 &\qquad
    0.614 \to 0.793
    \\
    0.246 \to 0.529 &\qquad
    0.464 \to 0.713 &\qquad
    0.590 \to 0.776
\end{array}\right).\label{Uij}
\eea
The range of experimental values of neutrino mass squared differences and leptonic mixing angles are given in Ref. \cite{PDG2015} as below
\bea
\sin ^{2}\theta _{12}&=&0.304 \pm  0.014, \hs \sin ^{2}\theta _{13}=(2.19\pm 0.12)\times 10^{-2},\crn
\De m_{21}^{2}&=&(7.53\pm 0.18)\times 10^{-5}\,\mathrm{eV^2},\crn
\sin ^{2}\theta _{23}&=&0.514^{+0.055}_{-0.056}\,\,  \mathrm{(normal \,\, mass \,\, hierarchy)},\crn
\sin ^{2}\theta _{23}&=&0.511\pm 0.055\,\,  \mathrm{(inverted \,\,  mass \,\,  hierarchy)},\crn
\De m_{32}^{2}&=& (2.44 \pm 0.06) \times 10^{-3}\,\mathrm{eV^2},  \mathrm{(normal \,\, mass \,\, hierarchy)},\crn
\De m_{32}^{2}&=& (2.49 \pm 0.06) \times 10^{-3}\,\mathrm{eV^2},  \mathrm{(inverted \,\, mass \,\, hierarchy)}.\label{PDG2015}
\eea
In fact, the models that successfully explain the experimental data are often mathematically complicate. An ideal physical model should be mathematically
quite simple but successfully explains the experimental data and its physical parameters can be tested by the experiments in near future. This desired model, up to now,
has not yet been effective because each model has its own advantages and disadvantages. To explain the specific neutrino mixings, it is simple to use discrete
symmetry such as $A_4, S_3, S_4$, etc.
The use of non-abelian discrete symmetries to construct the models describing the
lepton masses and mixings is a new method first proposed by E. Ma and G. Rajasekaran in 2001 \cite{A41}.
In this treatment, there are various models which have been proposed, see for example $A_4$ \cite{A41, A42, A43, A44, A45,A46, A47, A48, A49, A410, A411, A412, A413, A414,
 A415, A416, A417, A418, dlsh}, $S_3$\cite{S31,S32,S33,S34,S35,S36,S37,S38,S39,S310,S311,S312,S313,S314,S315,S316,S317,S318,S319,S320,S322,S323,S324,S325,S326,S327,S328,S329,S330,S331,S332,S333,S334,S335,S336,S337,S338,S339,S340,S341,S342}, $S_4$ \cite{S41,S42,S43,S44,S45,S46,S47,S48,S49,S410,S411,S413,S414,S415,S416,S417,S418,S419,S420,S421,S422,S423,S424,S425,S426,S427,S428,S429, S430},
$D_4$ \cite{D41,D42,D43,D45,D46,D47,D48,D49,D410,D411,D412}, $T'$ \cite{Tp1,Tp2,Tp3,Tp4,Tp7,Tp8,Tp9,Tp10,Tp11,Tp12}, $T_7$
\cite{T71, T72, T73, T74, T75}. However, in all above mentioned papers, the fermion masses and mixings generated from non-renormalizable interactions or at loop levels but not at tree-level.
 The models involving only renormalizable interactions were implemented in our previous works \cite{dlsvS4,dlnvS3,vlD4,vlS4,vlS3,vlT7, vD4, vT7, vlkS4, vlD4q, vlA4, alv27, vla27} in which the
discrete symmetries have been added to the 3-3-1 models. As we know the 3-3-1 model itself is an extension of the SM where the gauge group $SU(2)_L$ is extended to $SU(3)_L$.
In order to overcome such limitations, we studied a neutrino mass model by adding the discrete symmetry $S_4$ to the SM which accommodates the realistic lepton mass, mixing with non-zero
 $\theta_{13}$ and CP violation phase at the tree- level  with renormalizable interactions only \cite{VienS4SM}.

In  this paper, we construct a simple extension of the SM based on $T_7$ symmetry that leads to lepton mass, mixing with non-zero $\theta_{13}$ and CP violation phase\footnote{We note that $T_7$ symmetry has not been previously considered  in this kind of the model with the mentioned scenario. Furthermore, this model is different from our
previous works \cite{vlT7, vT7} because the 3-3-1 model (based on $SU(3)_C\otimes SU(3)_L\otimes U(1)_X$) itself is an extension of the SM.}. For this purpose, two $SU(2)_L$ doublets
and two $SU(2)_L$ singlets are introduced. The result follows
without perturbation and the number of scalars required to generate lepton masses are fewer than those in Ref.  \cite{VienS4SM}.

The future content of this paper reads as follows. In Sec.
\ref{leptonv} we present the fundamental elements of the model
and introduce necessary Higgs fields responsible for the
lepton masses. We summarize the results in the section \ref{conclus}. Finally, the appendices \ref{Solut1N1} and \ref{Solut1I1}  provide in detail solutions  for  neutrino masses in the normal and the inverted hierarchies, respectively.

\section{Lepton mass and mixing \label{leptonv}}

The lepton content of the model, under $\mathrm{SU}(2)_L\otimes
\mathrm{U}(1)_Y\otimes
\mathrm{U}(1)_X \otimes\underline{T}_7$ symmetries, is given in Tab. \ref{lepton}.
\begin{table}[h]
\caption{ \label{lepton}Lepton content of the model.}
{\begin{tabular}{@{}ccccccccc@{}} \toprule
             & \,\,$\psi_{L }$ &\, $l_{(1,2,3)R} $ &\,$\nu_{R} $ &\,\, $\phi $&\,\, $\varphi $&\,\, $\chi $&\,\, $\zeta $&\\
\noalign{\smallskip}\hline\noalign{\smallskip}
$\mathrm{SU}(2)_L$ &  $2$ &  $1$&$1$&\,\,$2$&\,\,$2$&\,\,$1$&\,\,$1$&\\
$\mathrm{U}(1)_Y$&  $-1$ & $-2$ &$0$&\,\,$1 $&\,\,$1$&\,\,$0$&\,\,$0$&  \\
$\mathrm{U}(1)_X$ & $1$  &$1$&$0$&   $0$  &\,\,$-1$& $0$  & $0$  \\
$\mathrm{T}_7$ &  $ \underline{3}$ &\, $\underline{1}, \underline{1}', \underline{1}''$ &\,\,$\underline{3}$&\,\,$\underline{3}$&\,\,$\underline{1}$&\,\,$\underline{3}$&\,\,$\underline{3}^*$& \\ \botrule
\end{tabular}}
\end{table}
The charged lepton masses arise from the couplings of
$\bar{\psi}_{L} l_{1R}, \bar{\psi}_{L} l_{2R}$ and $\bar{\psi}_{L}
l_{3R}$ to scalars, where $\bar{\psi}_{L} l_{iL}\, (i=1,2,3)$
transforms as $2$ under $\mathrm{SU}(2)_L$ and $\underline{3}^*$
under $T_7$. In order to generate masses for charged leptons, we need only
one $SU(2)_L$ Higgs doublets ($\phi$) lying in $\underline{3}$ under $T_7$, as given in Tab.\ref{lepton}.

The Yukawa interactions read
 \bea -\mathcal{L}_{l}&=&h_1 (\bar{\psi}_{L}\phi)_{\underline{1}} l_{1R}+
 h_2 (\bar{\psi}_{L}\phi)_{\underline{1}''}l_{2R}
+h_3 (\bar{\psi}_{i L}\phi)_{\underline{1}'} l_{3R}+H.c\crn
&=&h_1 (\bar{\psi}_{1L}\phi_1+\bar{\psi}_{2L}\phi_2+\bar{\psi}_{3L}\phi_3) l_{1R}\crn
&+&h_2 (\bar{\psi}_{1L}\phi_1+\om^2 \bar{\psi}_{2L}\phi_2+\om \bar{\psi}_{3L}\phi_3)l_{2R}\crn
&+&h_3 (\bar{\psi}_{1L}\phi_1+\om \bar{\psi}_{2L}\phi_2+\om^2 \bar{\psi}_{3L}\phi_3) l_{3R}+H.c. \label{clep}\eea
 In this work we
impose only the breaking $T_{7}\rightarrow Z_3$ in charged lepton sector, and this happens with the first alignment, i.e, $\langle \phi\rangle=
(\langle \phi_1\rangle,\langle \phi_1\rangle,\langle \phi_1\rangle )$ under $T_{7}$, where
\bea
\langle \phi_1\rangle=(0\hs v )^T.\label{vevphi}
\eea
With the  vacuum expectation value (VEV) of $\phi_1$ in Eq. (\ref{vevphi}), the mass Lagrangian for the charged leptons can be written in matrix form as
\bea
-\mathcal{L}^{\mathrm{mass}}_l=(\bar{l}_{1L},\bar{l}_{2L},\bar{l}_{3L})
M_l (l_{1R},l_{2R},l_{3R})^T+H.c,\eea
where \be M_l=
\left(%
\begin{array}{ccc}
  h_1v & h_2v & h_3 v \\
   h_1v &   \om^2h_2 v & \,\,\om h_3 v \\
  h_1v & \om h_2 v &\,\,\om^2 h_3 v \\
\end{array}%
\right).\label{Ml}\ee
The mass matrix $M_l$ in Eq. (\ref{Ml}) is  diagonalized :
 \bea U^\dagger_L M_lU_R=\left(%
\begin{array}{ccc}
  \sqrt{3}h_1 v & 0 & 0 \\
  0 & \sqrt{3}h_2 v& 0 \\
  0 & 0 & \sqrt{3}h_3 v\\
\end{array}%
\right)\equiv\left(%
\begin{array}{ccc}
  m_e & 0 & 0 \\
  0 & m_\mu & 0 \\
  0 & 0 & m_\tau \\
\end{array}%
\right),\eea
where
 \be
 m_e=\sqrt{3}h_1 v,\,\,
m_\mu= \sqrt{3} h_2 v,\,\, m_\tau=\sqrt{3} h_3 v, \label{memutau}
 \ee
and \bea U_L=\fr{1}{\sqrt{3}}\left(%
\begin{array}{ccc}
  1 & 1 & 1 \\
  1 & \om^2 & \om \\
  1 & \om & \om^2 \\
\end{array}%
\right),\hs U_R=1.\label{Ulep}\eea
The Yukawa couplings $h_{1, 2, 3}$ in charged
lepton sector are defined:
\bea h_1&=& \frac{m_e}{\sqrt{3}v},\hs  h_2= \frac{m_\mu}{\sqrt{3}v},\hs h_3= \frac{m_\tau}{\sqrt{3}v}.\label{h1h2h3}
 \eea
The experimental  values for masses of  the charged leptons are given in \cite{PDG2015}:
\bea m_e\simeq0.510998928\, \textrm{MeV},\hs \ m_{\mu}=105.6583715 \ \textrm{MeV},\hs m_{\tau}=1776.86\,
\textrm{MeV} \label{Lepmas}\eea
It follows that $h_1\ll h_2\ll h_3$. Furthermore, if we choose\footnote{In the SM, the
Higgs VEV $v$ is 246 GeV, fixed by the $W$ boson mass and the gauge coupling $m^2_W=\frac{g^2 }{4}v^2_{weak}$. In the model under consideration  $M^2_W\simeq\frac{3}{2}g^2v^2$. Therefore, we can identify
$v^2_{weak}=6v^2= (246 \, \mathrm{GeV})^2$ . It follows $ v\simeq 100 \, \mathrm{GeV}$.} the VEV $v \sim 100 \,\mathrm{GeV}$ then
\bea
h_1\sim 10^{-6},\,\,\, h_2\sim 10^{-4},\,\,\, h_3\sim 10^{-2},\label{hi}\eea
i.e, in the model under consideration, the hierarchy between the masses for charged-leptons can be achieved
 if there exists a hierarchy between Yukawa couplings $h_i \, (i=1,2,3)$ in
charged-lepton sector as given in Eq. (\ref{hi}). We note that the masses of charged  leptons are self-separated by only one
Higgs triplet $\phi$ (the same as in the SM), and this is a good feature
 of the $T_7$ group. We remind that the models with the other discrete symmetry groups need more than one Higgs scalar in the charged lepton sector.

The neutrino masses arise from the couplings of $\bar{\psi}_{L} \nu_{R}$
and $\bar{\nu}^c_{R} \nu_{R}$ to scalars, where  $\bar{\psi}_{L} \nu_{R}$ transforms as
$2$ under $\mathrm{SU}(2)_L$ and $\underline{1}\oplus\underline{1}'\oplus\underline{1}''\oplus
\underline{3}\oplus \underline{3}^*$ under $T_7$;
$\bar{\nu}^c_{R} \nu_{R}$ transform as $1$ under $\mathrm{SU}(2)_L$
and $\underline{3}\oplus \underline{3}^*\oplus \underline{3}^*$
under $T_7$. Note that $\underline{3} \otimes \underline{3} \otimes \underline{3}$ has two
invariants and $\underline{3} \otimes \underline{3} \otimes \underline{3}^*$
has one invariant under $T_7$. In order to generate mass for neutrinos, we additionally introduce
one $SU(2)_L$ doublet $(\varphi)$ and two $SU(2)_L$ singlets $(\chi,\, \zeta)$, respectively, put in $\underline{1}$, $\underline{3}$
and $\underline{3}^*$ under $T_7$ as given in Tab. \ref{lepton}. We note that the $U(1)_X$ symmetry
 forbids the Yukawa terms of the form $(\bar{\psi}_L \tilde{\phi})_{\underline{3}_s}\nu_{R}$ and yield the expected results in neutrino sector,
 and this is interesting feature of $X$-symmetry.  It is also interesting to note that $\varphi$ contributes to the Dirac mass matrix, $\chi $ and $\zeta$ contribute
 to the Majorana mass matrix of the right-handed neutrinos.
 In fact, there exist no one-dimensional representation in $\underline{3}\otimes\underline{3}$ under $T_{7}$. Hence, $\zeta$ put in $\underline{3}^*$ will be responsible
 for a realistic neutrino spectrum without any perturbation and soft breaking in both lepton  and neutrino sectors.
 This feature is different from the one in Ref. \cite{A42015}.
It needs to note that $\varphi$ contributes to the Dirac mass matrix in
the neutrino sector, $\chi$  and $\zeta$ contribute to the Majorana mass matrix of the right-handed neutrinos. The interesting feature of $X$-symmetry is to prevents the
unwanted interaction of the form $(\bar{\psi}_L \tilde{\phi})_{\underline{3}_s}\nu_{R}$ and provides the expected results in the neutrino sector.

In this work we impose that the breaking $T_{7}\rightarrow  \{\mathrm{identity}\}$ must be taken place, i.e, $T_{7}$ is completely broken in neutrino sector.
This can be achieved within each case below.
 \ben
\item [(1)] A new $\mathrm{SU}(2)_L$ singlet $\chi$ lies in $\underline{3}$ under $T_{7}$ with the VEV is given by $\langle \chi \rangle=(0, \langle \chi_2\rangle,0)^T$ under $T_{7}$, where
\be
\langle \chi_2\rangle=v_\chi. \label{chivev}\ee
\item [(2)] Another
singlet $\zeta$ lies in $\underline{3}^*$ under $T_{7}$  with the VEV is given by $\langle \zeta\rangle=(\langle \zeta_1,
\langle \zeta_2\rangle, \langle \zeta_3\rangle)^T$ under $T_{7}$, i.e. $\langle \zeta_1\rangle \neq \langle \zeta_2\rangle\neq \langle \zeta_3\rangle\neq 0$, where
\bea \langle \zeta_i\rangle=u_i \hs (i=1,2,3). \label{vzetai} \eea \een

The neutrino Yukawa interactions are given by
\bea
 -\mathcal{L}_{\nu}&=&x (\bar{\psi}_L \tilde{\varphi})_{{3}^*}\nu_{R}+\fr y 2  (\bar{\nu}^c_R\chi)_{{3}^*}\nu_{R}+
\fr z 2  (\bar{\nu}^c_R\zeta)_{{3}^*}\nu_{R}+H.c\crn
 &=& x(\bar{\psi}_{1L}\tilde{\varphi}\nu_{1R}+\bar{\psi}_{2L}\tilde{\varphi}\nu_{2R}+\bar{\psi}_{3L}\tilde{\varphi}\nu_{3R})\crn
&+&\frac{y}{2}\left[(\bar{\nu}^c_{2R}\chi_{3}+\bar{\nu}^c_{3R}\chi_{2})\nu_{1R}
+(\bar{\nu}^c_{3R}\chi_{1}+\bar{\nu}^c_{1R}\chi_{3})\nu_{2R}
+(\bar{\nu}^c_{1R}\chi_{2}+\bar{\nu}^c_{2R}\chi_{1})\nu_{3R}\right]\crn
&+& \frac{z}{2}(\bar{\nu}^c_{1R}\zeta_2\nu_{1R}+\bar{\nu}^c_{2R}\zeta_3\nu_{2R}+\bar{\nu}^c_{3R}\zeta_1\nu_{3R})+H.c.\label{Lny}\eea
The neutrino mass Lagrangian are given as
\bea -\mathcal{L}^{mass}_{\nu}&=&xv(\bar{\nu}_{1L}\nu_{1R}+ \bar{\nu}_{2L}\nu_{2R}+\bar{\nu}_{3L}\nu_{3R})\crn
&+&\frac{y}{2}\left(v_\chi \bar{\nu}^c_{3R}\nu_{1R}
+v_\chi\bar{\nu}^c_{1R}\nu_{3R}+v_\chi \bar{\nu}^c_{2R}\nu_{1R}
+v_\chi\bar{\nu}^c_{1R}\nu_{2R}\right)\crn
&+& \frac{z}{2}(u_2\bar{\nu}^c_{1R}\nu_{1R}+u_3\bar{\nu}^c_{2R}\nu_{2R}+u_1\bar{\nu}^c_{3R}\nu_{3R})+H.c.\label{Ynumass}
 \eea
We can rewrite in the matrix form \bea -\mathcal{L}^{\mathrm{mass}}_\nu&=&\fr 1 2
\bar{\chi}^c_L M_\nu \chi_L+ H.c.,\hs  \chi_L\equiv
\left(%
\begin{array}{c}
  \nu^c_L \\
  \nu_R \\
\end{array}%
\right),\hs M_\nu\equiv\left(%
\begin{array}{cc}
  0 & M_D \\
  M^T_D & M_R \\
\end{array}%
\right), \label{MnuLDR}\\
 \nu^c_L&=&(\nu^c_{1L}\hs\nu^c_{2L}\hs\nu^c_{3L})^T,\hs
\nu_R=(\nu_{1R}\hs \nu_{2R}\hs\nu_{3R})^T, \nn  \eea where the Dirac neutrino mass
matrix ($M_{D}$) and the right-handed Majorana neutrino mass
matrix $(M_{R})$ are given by
\bea M_D &=&
 \left(%
\begin{array}{ccc}
a &0 & 0 \\
0 &a & 0 \\
0 &0 & a\\
\end{array}%
\right),\,\,
M_R=
\left(%
\begin{array}{ccc}
 N_2 & 0 & b \\
 0 &N_3 & 0 \\
 b &0 & N_1 \\
\end{array}%
\right),  \label{MLDR}\eea with
\bea
a&=& v_\varphi x,\hs b=v_\chi y,\hs N_i =u_i z \hs (i=1,2,3).\label{abNi}
\eea
The seesaw mechanism generates small masses for neutrinos is given by
 \bea
M_{\mathrm{eff}}=-M_DM_R^{-1}M^T_D=\left(%
\begin{array}{ccc}
  A_1 & 0 & B \\
  0 & A_3 & 0 \\
  B & 0 & A_2 \\
\end{array}%
\right), \label{Mef}\eea
where
\bea
A_1&=&\frac{a^2  N_1}{ b^2-N_1N_2}, \hs A_2=\frac{a^2 N_2}{ b^2-N_1N_2},\hs
A_3=-\frac{a^2}{N_3},\hs B =\frac{a^2b}{N_1 N_2-b^2}.\label{A123B}
\eea
The matrix $M_{eff}$ in Eq. (\ref{Mef}) has three exact eigenvalues given by
\bea m_1&=&\fr 1 2 \left(A_1 + A_2 -\sqrt{(A_1-A_2)^2+4 B^2}\right), \hs m_2=A_3,\crn
 m_{3} &=&\fr 1 2 \left(A_1 + A_2 +\sqrt{(A_1-A_2)^2+4 B^2}\right),\label{m123}\eea
and the corresponding eigenstates are
 \bea U_\nu&=&\left(%
\begin{array}{ccc}
 \fr{K}{\sqrt{K^2+1}}& 0 & \fr{1}{\sqrt{K^2+1}} \\
0 & 1 & 0 \\
\fr{1}{\sqrt{K^2+1}} & 0 & -\fr{K}{\sqrt{K^2+1}} \\
\end{array}%
\right)\, ,\label{Unu1}\eea
where
\be
K=\frac{A_1 -A_2 -\sqrt{(A_1 -A_2)^2+4 B^2}}{2B},\label{K}
\ee
and $A_{1,2}, B$ are given in Eq. (\ref{A123B}).

The lepton mixing matrix is then expressed as
\bea U=U^\dagger_L
U_\nu= \fr{1}{\sqrt{3}}\left(%
\begin{array}{ccc}
  \fr{1+K}{\sqrt{K^2+1}} & 1 &  \fr{1-K}{\sqrt{K^2+1}} \\
\fr{K+\om^2}{\sqrt{K^2+1}} & \om &  \fr{1-K\om^2}{\sqrt{K^2+1}} \\
\fr{K+\om}{\sqrt{K^2+1}}    & \om^2 &  \fr{1-K\om}{\sqrt{K^2+1}} \\
\end{array}%
\right)\, ,\label{Ulep}\eea
where $K$ is defined in Eq.(\ref{K}).
Comparing  the lepton mixing matrix in Eq. (\ref{Ulep}) and  the standard parametrization
\footnote{In fact, the Majorana phases do not contribute to neutrino
oscillations so they will be ignored for the rest of this work.} in Ref. \cite{PDG2015} yields:
 \bea s_{13} e^{-i \delta}&=&\fr{1}{\sqrt{3}}\fr{1-K}{\sqrt{K^2+1}},\label{s13}\\
 t^2_{12}&=&\left|\frac{\sqrt{K^2+1}}{1+K}\right|^2,\label{t12}\\
 t^2_{23}&=&\left|\frac{1-K\om^2}{1-K\om}\right|^2.\label{t23}\eea

In the case $K$ being real numbers, Eqs. (\ref{s13}) and (\ref{t23}) imply $ \theta_{23} =45^o$ and $\delta=0$. As we know, the recent experimental data
imply $\delta\neq 0$. To overcome this, we will consider $K$ as a complex variable.
Substituting $\om =-\frac{1}{2}+i\frac{\sqrt{3}}{2}$ into Eqs. (\ref{s13}),  (\ref{t12}) and  (\ref{t23}) we obtain:
\bea s_{13}&=&\frac{1}{\sqrt{3}}\frac{\left[(k_1-1)^2+k_2^2\right]^{1/2}}{\al^{1/4}},\label{s131}\\
t_{12}^2&=&\frac{\al^{1/2}}{(1+k_1)^2+k_2^2},\label{t121}\\
t^2_{23}&=&1-\frac{2\sqrt{3}k_2}{1+k_1+k^2_1+k^2_2+\sqrt{3}k_2},\label{t231}
\eea
where
\be
\al=(1+k_1^2-k_2^2)^2+4k_1^2k_2^2, \label{alpha}
\ee
and $k_1$ and $k_2$  being the real and imaginary parts of $K$, respectively.

On the other hand, from Eq.(\ref{s13}), we get:
\bea  e^{-i \delta}&=&\fr{1}{s_{13}\sqrt{3}}\fr{1-K}{\sqrt{K^2+1}}\equiv \cos \delta-i\sin \delta,\label{e}\eea
with
\bea
 \cos \delta&=&\left(1+2k_1-k^2_1-k^2_2-\sqrt{\al}\right)\beta,\crn
  \sin \delta &=&\left\{k^2_2-1 + k_1(1 - k_1 + k_1^2 +k_2^2) + (1 - k_1)\sqrt{\al}\right\}\beta, \label{sincos}
\eea
where
\bea
 \beta&=&\frac{\al^{1/4}\sqrt{-1-k^2_1+k^2_2+\sqrt{\al}}}
 {\sqrt{2}\sqrt{(k_1-1)^2+k^2_2}\left[(-1-k^2_1+k^2_2)\sqrt{\al}+k^4_1+(k^2_2-1)^2+2k^2_2(1+k^2_2)\right]},\label{beat}
\eea
which is satisfying the relation $\sin^2\delta+\cos^2\delta =1$ with all $k_1, k_2$.

The neutrino mass spectrum can be the normal hierarchy ($
|m_1|\simeq |m_2| < |m_3|$), the inverted hierarchy ($|m_3|< |m_1|\simeq |m_2|$)
 or nearly degenerate ($|m_1|\simeq |m_2|\simeq |m_3| $). The mass
ordering of neutrino depends on the sign of $\Delta m^2_{23}$
which is currently unknown. However, some tight upper limits on the total neutrino mass $\sum m_\nu$ have given by the recent studies. For example, the
total mass of three degenerate neutrinos was given by Planck satellite mission \cite{Ade2015}, $\sum m_\nu<0.72~\textrm{eV}$ ($95\%$ CL) by using Planck
TT+lowP data, and $\sum m_\nu<0.49~\textrm{eV}$ ($95\%$ CL) by using Planck TT,TE,EE+lowP data. While the improved constraints are given by adding the baryon
 acoustic oscillation (BAO) measurements \cite{SWang}, i.e., $\sum m_\nu<0.21~\textrm{eV}$ ($95\%$ CL) and $\sum m_\nu<0.17~\textrm{eV}$ ($95\%$ CL), respectively.
 Another upper limit  was given in Ref. \cite{Zhang2015}, $\sum m_\nu<0.113~\textrm{eV}$ ($95\%$ CL).

As will see, in
the model under consideration,  the two possible signs of $\Delta
m^2_{23}$ correspond to two types of neutrino mass spectrum as well as the values
of the atmospheric neutrino mixing angle $\theta_{23}$ can be
provided.

Combining Eq. (\ref{s131})  with the experimental values of $\theta_{13}$ given in Ref. \cite{PDG2015}
as shown in Eq.(\ref{PDG2015}), we have a solution as follow\footnote{There exist four mathematical solutions, however, these solutions differ only by the sign
of  $m_{1,2,3}$ which has no effect on the neutrino oscillation experiments.}:
\bea
k_2=-\frac{1}{2}\sqrt{(8.03468 - 4 k_1) k_1-4.03468 +2\sqrt{0.069663 + (0.139025 k_1-0.139326) k_1}}.\label{k2}
\eea

Next,  from Eqs. (\ref{k2}) and (\ref{t231}) with the experimental values of $\theta_{23}$ in Eq.(\ref{PDG2015}), we get two solutions\footnote{Here we only
consider one case because another value has no effect on the neutrino oscillation experiments.}:
\bea
k_1=0.690532, \hs k_2=-0.0350532,\hs K=0.690532-0.0350532i, \label{k1k2}
\eea
and the lepton mixing matrix in (\ref{Ulep}) then takes the form
\bea \left|U\right|\simeq\left(%
\begin{array}{ccc}
 0.803441 &\hs 0.57735 &\hs 0.147986\\
0.437621 &\hs 0.57735 &\hs 0.709451 \\
0.405089 &\hs 0.57735 &\hs 0.689859 \\
\end{array}%
\right),\label{Ulepmix1}\eea
which is consistent with constraint in Eq.(\ref{Uij}).
Now, substituting $k_{1,2}$ from (\ref{k1k2}) in to (\ref{t121}) yields\footnote{$\theta_{12}\simeq 35.7^\circ$ obtained from the model is an acceptable
 prediction.} $t^2_{12}=0.516381$ (or $t_{12}=0.718597$), i.e, $\theta_{12}\simeq 35.7^\circ$. It follows $\cos\delta = -0.991667,\, \sin\delta= 0.128827$, i.e, $\delta \simeq 172.598^\circ$.

Combining (\ref{K}) and the values of $K$ in
(\ref{k1k2}), we obtain
\bea A_1 = A_2 - (0.753905+ 0.108377i) B. \label{A1A2B} \eea

\subsection{Normal spectrum ($\Delta m^2_{23}> 0$)}
Substituting $A_1$ from
(\ref{A1A2B}) into (\ref{m123}) and combining with the two
experimental constraints on squared mass differences of neutrinos for the normal spectrum
as shown in (\ref{PDG2015}), i.e, $\De m_{21}^{2}=7.53\times 10^{-5}\,\mathrm{eV^2},\,\, \De m_{32}^{2}= 2.44  \times 10^{-3}\,\mathrm{eV^2}$,
we get the analytical expressions of $A_2, B, m_{1,2,3}$ (in [eV]) given
in Appendix \ref{Solut1N1}.

\begin{figure}[ht]
\bc
\includegraphics[width=13.0cm, height=5.0cm]{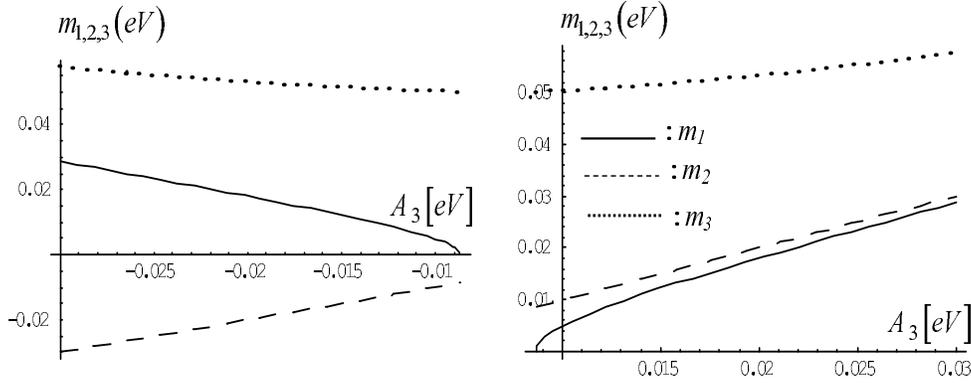}
\vspace*{-0.1cm} \caption[$m_{1,2,3}$ as functions of $A_3$ in the Normal spectrum with
 $A_3\in(-0.03, -0.0087) \, \mathrm{eV}$ (left) and $A_3\in(0.0087, 0.03) \, \mathrm{eV}$ (right).]{$m_{1,2,3}$ as functions of $A_3$ in the normal spectrum with
 $A_3\in(-0.03, -0.0087) \, \mathrm{eV}$ (left) and $A_3\in(0.0087, 0.03) \, \mathrm{eV}$ (right).}\label{m1231N1}
\ec
\end{figure}

\begin{figure}[ht]
\begin{center}
\includegraphics[width=6.0cm, height=4.5cm]{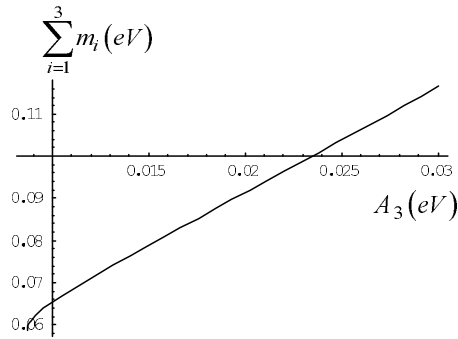}
\vspace*{-0.1cm}
\caption[The sum $\sum^3_{i=1}m_i$ as a function of
$A_3$ with $A_3 \in (0.0087, 0.03)\,\mathrm{eV}$ in the
normal spectrum.]{The sum $\sum^3_{i=1}m_i$ as a function of
$A_3$ with $A_3 \in (0.0087, 0.03)\,\mathrm{eV}$ in the
normal spectrum.}\label{m123sN1}
\end{center}
\end{figure}

By using the upper limits on neutrino mass   \cite{Ade2015, SWang, Zhang2015} we can
 restrict $A_3 \leq 0.72\,\mathrm{eV}$. However, in the normal spectrum case in (\ref{Solut1N1}),
  $A_3 \in [0.0087, 0.03]\, \mathrm{eV}$ or $A_3 \in [-0.03, -0.0087]\, \mathrm{eV}$ are good regions of $A_3$
that can reach the realistic neutrino mass hierarchies. With  $m_2 \in [0.0087, 0.03]\, \mathrm{eV}$, $m_{1,2,3}$
as functions of $A_3=m_2$ are plotted in Fig.\ref{m1231N1} . This figure shows
  that there exist allowed regions of the parameter $A_3$ where either normal
  or quasi-degenerate neutrino masses spectrum is achieved.
  The quasi-degenerate mass hierarchy\footnote{There is no clear limits between neutrino mass hierarchies by the
   recent experimental results on neutrino oscillations} is obtained when $A_3\in
 [0.03\,\mathrm{eV} , +\infty$) or  $A_3\in
 (-\infty, -0.03\,\mathrm{eV} $] ($|A_3|$ increases
  but must be small enough because of the scale of $m_{1,2,3}$). The normal
  mass hierarchy will be obtained if $A_3\in [0.0087, 0.03]\,
   \mathrm{eV}$ or $A_3\in [-0.03, -0.0087]\, \mathrm{eV}$.
The total neutrino masses in the model under consideration $\sum^3_{i=1}m_i$ and $\sum^3_{i=1}|m_i|$ with
$m_2 \in [0.0087, 0.05]\,\mathrm{eV}$ is depicted in Fig.\ref{m123sN1}.

\begin{figure}[ht]
\begin{center}
\includegraphics[width=6.5cm, height=4.5cm]{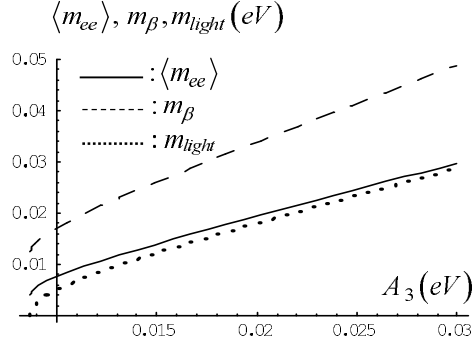}
\vspace*{-0.1cm} \caption[ $m_{ee}$, $m_{\beta}$ and $m_{light}$ as functions of $A_3$
with $A_3\in(0.00867, 0.05) \,
\mathrm{eV}$ in the normal spectrum.]{$m_{ee}$, $m_{\beta}$ and $m_{light}$ as functions of $A_3$
with $A_3\in(0.00867, 0.05) \,
\mathrm{eV}$ in the normal spectrum.}\label{mee1T7}
\vspace*{-0.3cm}
\end{center}
\end{figure}

It is easily to obtain the effective mass $\langle m_{ee}\rangle$ governing neutrinoless double beta decay
 \cite{betdecay1, betdecay2,betdecay3,betdecay4,betdecay5, betdecay6} $\langle m_{ee}\rangle=  \left|\sum^3_{i=1} U_{ei}^2 m_i \right|$, $
m_{\beta} = \left\{\sum^3_{i=1} |U_{ei}|^2 m_i^2 \right\}^{1/2}$ by combining  the expressions (\ref{Ulep}), (\ref{k1k2}), (\ref{m1N1}), (\ref{m2N1})
 and (\ref{m3N1}), the values of $m_{ee}, m_{\beta}$ and  $m_{light}$ are plotted in Fig.\ref{mee1T7} together with  $ m_1$ with $A_3 \in (0.0087, 0.03)\,\mathrm{eV}$. We also note that in the normal spectrum, $m_1\approx m_2<m_3$ so $m_{light}= m_1$
given in (\ref{m2N1}) is the lightest neutrino mass.

To get explicit values of the model
parameters, we assume $A_3\equiv m_2=10^{-2}\, \mathrm{eV}$, which is safely small.
Then the other neutrino masses and the other parameters are explicitly
 given in Tab. \ref{para1}.

  \begin{table}[ht]
\caption{ \label{para1} The model parameters in the case $A_3 =10^{-2}\, \mathrm{eV}$  in the normal spectrum}
{\begin{tabular}{@{}cccccccccc@{}} \toprule
& Parameters $[ \mathrm{eV}]$&\hs The derived values&\\\hline
& \,\,$A_1$ &\, $0.0357232+0.00100894i$&\\
 & \,\,$A_2$ &\, $0.0196452-0.00100894i$&\\
  & \,\,$B$ &\, $-0.0212715+0.000381301i$&\\
   & \,\,$m_1$ &\, $0.00496991$&\\
    & \,\,$m_3$ &\, $0.0503984$&\\
     & \,\,$\sum m^I_{i} $ &\, $0.0653683$&\\
      & \,\,$\langle m^I_{ee}\rangle$ &\, $0.00761271$&\\
       & \,\,$m^I_{\beta}$ &\, $0.0171627$&\\
\botrule
\end{tabular}}
\end{table}

Now, comparing Eqs.  (\ref{A123B}) and derived values in Tab. \ref{para1} we get the relations:
\bea
N_1&=&-(142.621 + 4.02808i)a^2, \hs N_2=(-78.4315+4.02808i)a^2, \crn
N_3&=&-100 a^2, \hs b=(-84.9243+1.52231 i)a^2.\label{relat1N}
\eea
or
\bea
|N_1|/|b|&=&1.67979,\,\,\, |N_2|/|b| =0.924614,\,\,\, |N_3|/|b|=1.17733,\label{relat2N}\\
|N_1/a^2|&=&142.678,\,\,\, |N_2/a^2| =78.5348, \,\,\, |N_3/a^2|=100,\,\,\, |b/a^2|=84.938,\label{relat3N}
\eea
i.e.,  $N_1$, $N_2$, $N_3$ and $b$ have the same order of magnitude, and approximately  two orders of magnitude of $a^2$.

\subsection{Inverted spectrum ($\Delta m^2_{23}< 0$)}
Substituting $A_1$ in (\ref{A1A2B}) into (\ref{m123}) and combining with the experimental constraints on squared mass differences of neutrinos for the inverted spectrum
as shown in (\ref{PDG2015}), i.e, $\De m_{21}^{2}=7.53\times 10^{-5}\,\mathrm{eV^2},\,\, \De m_{32}^{2}= -2.49  \times 10^{-3}\,\mathrm{eV^2}$, we get a solution (in [eV]) given in Appendix \ref{Solut1I1}.

\begin{figure}[ht]
\bc
\includegraphics[width=13.0cm, height=5.0cm]{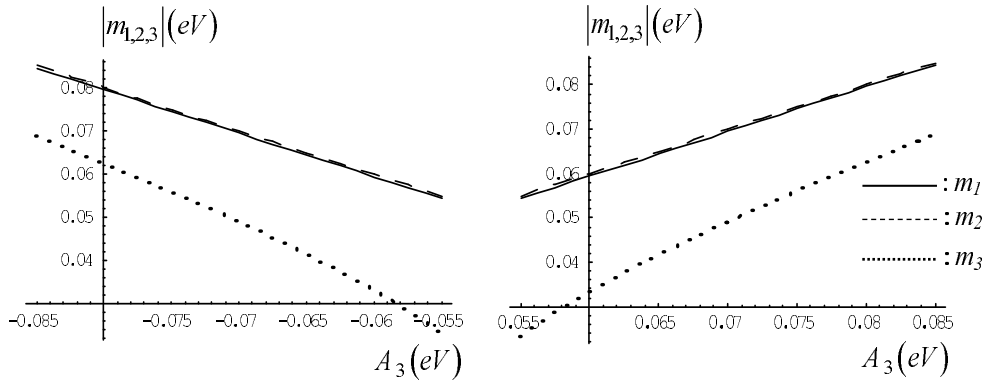}
\vspace*{-0.1cm} \caption[$|m_{1,2,3}|$ as functions of $A_3$ in the inverted spectrum with
 $A_3\in(-0.085, -0.055) \, \mathrm{eV}$ (left) and $A_3\in(0.055, 0.085) \, \mathrm{eV}$ (right).]{$|m_{1,2,3}|$ as functions of $A_3$ in the inverted spectrum with
 $A_3\in(-0.085, -0.055) \, \mathrm{eV}$ (left) and $A_3\in(0.055, 0.085) \, \mathrm{eV}$ (right).}\label{m1231I1}
\ec
\end{figure}

\begin{figure}[ht]
\begin{center}
\includegraphics[width=6.5cm, height=5.0cm]{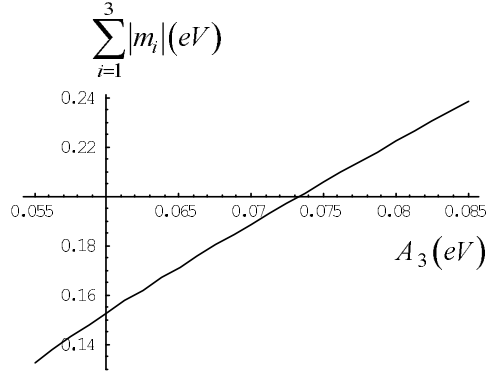}
\vspace*{-0.1cm}
\caption[The sum $\sum^3_{i=1}m^I_i$ as a function of
$A_3$ with $A_3 \in (0.055, 0.085)\,\mathrm{eV}$ in the
inverted spectrum.]{The sum $\sum^3_{i=1}m^I_i$ as a function of
$A_3$ with $A_3 \in (0.055, 0.085)\,\mathrm{eV}$ in the
inverted spectrum.}\label{m123sI1}
\end{center}
\end{figure}

In the inverted spectrum, with the solution  in (\ref{Solut1I1}),
  $A_3 \in (0.055, 0.085)\, \mathrm{eV}$ or $A_3 \in [-0.085, -0.055]\, \mathrm{eV}$ are good regions of $A_3$
that can reach the inverted neutrino mass hierarchies. The absolute values $|m_{1,2,3}|$
as functions of $A_3=m_2$ are plotted in Fig. \ref{m1231I1} in which  $A_3 \in [0.055, 0.085]\, \mathrm{eV}$. This figure shows
  that the quasi-degenerate mass hierarchy is obtained when $A_3\in
 [0.085\,\mathrm{eV} , +\infty$) or  $A_3\in
 (-\infty, -0.085\,\mathrm{eV} $] . The inverted  mass hierarchy will be obtained if $|A_3|\in [0.055, 0.085]\,
   \mathrm{eV}$. The total neutrino masses $\sum^3_{i=1}m^I_i$ and $\sum^3_{i=1}|m^I_i|$ with $A_3 \in [0.055, 0.085]\,\mathrm{eV}$ is depicted in Fig.\ref{m123sI1}.

\begin{figure}[ht]
\begin{center}
\includegraphics[width=7.0cm, height=5.5cm]{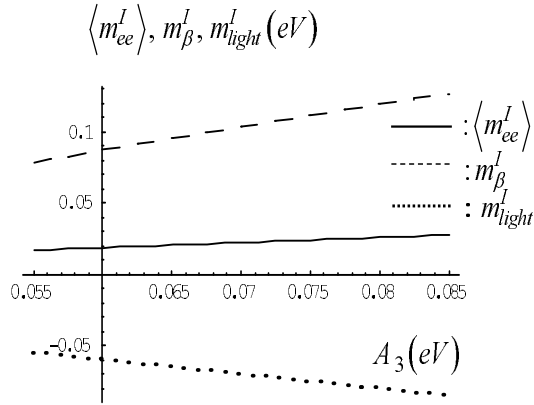}
\vspace*{-0.1cm} \caption[ $m_{ee}$, $m_{\beta}$ and $m_{light}$ as functions of $A_3$
with $A_3\in (0.055, 0.085) \,
\mathrm{eV}$ in the inverted spectrum.]{$m_{ee}$, $m_{\beta}$ and $m_{light}$ as functions of $A_3$
with $A_3\in (0.055, 0.085) \,
\mathrm{eV}$ in the inverted spectrum.}\label{mee1I}
\vspace*{-0.3cm}
\end{center}
\end{figure}

In the inverted spectrum, the effective mass $\langle m^I_{ee}\rangle$ governing neutrinoless double beta decay $\langle m^I_{ee}\rangle$
and $m^I_\beta $ together with $m_3$ are plotted in Fig.\ref{mee1I} with
$A_3 \in [0.055, 0.085]\,\mathrm{eV}$ by combining the expressions (\ref{Ulep}) , (\ref{k1k2}), (\ref{m1I1}), (\ref{m2I1}) and (\ref{m3I1}).
 In this case $m^I_{light}= m_3$
given in Eq. (\ref{m3N1}) is the lightest neutrino mass.

To get explicit values of the model
parameters, we assume $A_3\equiv m_2=6\times10^{-2}\, \mathrm{eV}$. The other neutrino masses and the other parameters are explicitly
 given in Tab. \ref{para2}.
  \begin{table}[h]
\caption{ \label{para2} The model parameters in the case $A_3 =6\times10^{-2}\, \mathrm{eV}$  in the inverted spectrum}
{\begin{tabular}{@{}cccccccccc@{}} \toprule
& Parameters $[ \mathrm{eV}]$&\hs The derived values &\\\hline
& \,\,$A_1$ &\, $-0.0417327+0.000578609i$&\\
 & \,\,$A_2$ &\, $-0.0509532-0.000578609 i$&\\
  & \,\,$B$ &\, $-0.0121988+0.00021867i$&\\
   & \,\,$m_1$ &\, $-0.0593692$&\\
    & \,\,$m_3$ &\, $-0.0333167$&\\
     & \,\,$\sum m^I_{i} $ &\, $0.0326858$&\\
      & \,\,$\langle m^I_{ee}\rangle$ &\, $0.0190284$&\\
       & \,\,$m_{\beta}$ &\, $0.08723$&\\
\botrule
\end{tabular}}
\end{table}

Comparing Eqs.  (\ref{A123B}) and derived values in Tab. \ref{para2} yields:
\bea
N_1&=&(21.0986-0.292525i)a^2, \hs N_2=(25.7602+0.292525i)a^2, \crn
N_3&=&-16.6667 a^2, \hs b=(-6.16732+0.110552 i)a^2.\label{relat1I}
\eea
or
\bea
|N_1|/|b|&=&3.42082,\,\,\, |N_2|/|b| =4.17648,\,\,\, |N_3|/|b|=2.70198,\label{relat2I}\\
|N_1/a^2|&=&21.1006,\,\,\, |N_2/a^2| =25.7618, \,\,\, |N_3/a^2|=16.6667,\,\,\, |b/a^2|=6.16831,\label{relat3I}
\eea
i.e., $N_1$, $N_2$, $N_3$ and $b$ have the same order of magnitude, and approximately  one orders of magnitude of $a^2$.

 \section{\label{conclus}Conclusions}
 We have proposed a simple Standard Model extension based on $T_7$ flavor symmetry
 which  accommodates lepton mass, mixing with non-zero $\theta_{13}$ and
 CP violation phase. The spontaneous symmetry breaking in the model is imposed to
 obtain the realistic lepton mass and mixing pattern at the tree- level
 with renormalizable interactions. In difference from other discrete groups, the $T_7$ flavor group
 requires only one VEV ($ \langle \phi_1 \rangle = v$) which, the same as in the SM, is enough for production of the charged lepton masses.
 The neutrinos get small masses from
 one $SU(2)_L$ doublet and two $SU(2)_L$ singlets  in which one being in
  $\underline{1}$ and the two others in $\underline{3}$ and $\underline{3}^*$ under $T_7$, respectively.
  The model also gives a remarkable prediction
  of Dirac CP violation $\delta_{CP}=172.598^\circ$ in both normal and inverted spectrum
  which is still missing in the neutrino mixing matrix.

\section*{Acknowledgments}
This research has received funding from the Vietnam
National Foundation for Science and Technology Development
(NAFOSTED) under Grant number 103.01-2014.51.

\appendix
\section{\label{Solut1N1} Neutrino masses  in the normal spectrum}
\bea
A_2&=&8.44732\times 10^{-9}\sqrt{\Ga}+(8.13873\times 10^{-6}+2.91875\times 10^{-7}i)\sqrt{\ga.\Ga}\crn
&-&(0.188476+0.0270942i)\sqrt{\ga'-2\sqrt{\ga}}+7.14452\times 10^{-6}A^2_3\sqrt{\Ga},\crn
B&=&-0.5\sqrt{\ga'-2\sqrt{\ga}},\crn
m_1&=&-0.5\sqrt{0.0023647+2A^2_3-(2.27831+0.081706 i)\sqrt{\ga}}\crn
&+&(0.188476+0.0270943i)\sqrt{\ga'-2\sqrt{\ga}}+8.44732\times 10^{-9}\sqrt{\Ga}\crn
&+&(8.13873\times 10^{-6}+2.91875\times 10^{-7}i)\sqrt{\ga.\Ga}\crn
&-&(0.188476+0.0270942i)\sqrt{\ga'-2\sqrt{\ga}}+7.14452A^2_3\sqrt{\Ga}, \label{m1N1}\\
m_2&=&A_3, \label{m2N1}\\
m_3&=&0.5\sqrt{0.0023647+2A^2_3-(2.27831+0.081706 i)\sqrt{\ga}}\crn
&+&(0.188476+0.0270943i)\sqrt{\ga'-2\sqrt{\ga}}+8.44732\times 10^{-9}\sqrt{\Ga}\crn
&+&(8.13873\times 10^{-6}+2.91875\times 10^{-7}i)\sqrt{\ga.\Ga}\crn
&-&(0.188476+0.0270942i)\sqrt{\ga'-2\sqrt{\ga}}+7.14452\times 10^{-6}A^2_3\sqrt{\Ga}.\label{m3N1}
\eea
where
\bea
\ga &=&(-1.4104\times 10^{-7}+1.01291\times 10^{-8}i)+(0.00181524-0.000130366i)A^2_3\crn
&+&(0.76764-0.0551299i)A^4_3,\label{ga}\\
\ga'&=&(0.00207317-0.0000743489i)+(1.75343-0.0628823i)A^2_3,\label{gap}\\
\Ga &=&(7.32235\times 10^{12}+0.000183105i)+(6.19305\times 10^{15}-0.0625i)A_3\crn
&-&(7.05485\times 10^{15} +2.53004\times 10^{14}i)\sqrt{\ga}.\label{Ga}
\eea
\section{\label{Solut1I1} Neutrino masses  in the inverted spectrum}
\bea
A_2&=&9.5457\times 10^{-9}\sqrt{\Ga_1}-(8.4778\times 10^{-6}+3.04035\times 10^{-7}i)\sqrt{\ga_1.\Ga_1}\crn
&-&(0.188476+0.0270942i)\sqrt{\ga'_1-2\sqrt{\ga_1}}-7.44217\times 10^{-6}A^2_3\sqrt{\Ga_1},\crn
B&=&-0.5\sqrt{\ga'_1-2\sqrt{\ga_1}},\crn
m_1&=&-0.5\sqrt{(-0.0025653+2.71051\times 10^{-20}i)+2A^2_3-(2.27831+0.081706 i)\sqrt{\ga_1}}\crn
&+&(0.188476+0.0270942i)\sqrt{\ga'_1-2\sqrt{\ga_1}}
-(8.4778\times 10^{-6}+3.04035\times 10^{-7}i)\sqrt{\ga_1.\Ga_1}
\crn
&+&(9.5457\times 10^{-9}-7.44217\times 10^{-6}A^2_3)\sqrt{\Ga_1}, \label{m1I1}\\
m_2&=&A_3, \label{m2I1}\\
m_3&=&0.5\sqrt{(-0.0025653+2.71051\times 10^{-20}i)+2A^2_3-(2.27831+0.081706 i)\sqrt{\ga_1}}\crn
&+&(0.188476+0.0270942i)\sqrt{\ga'_1-2\sqrt{\ga_1}}
-(8.4778\times 10^{-6}+3.04035\times 10^{-7}i)\sqrt{\ga_1.\Ga_1}
\crn
&+&(9.5457\times 10^{-9}-7.44217\times 10^{-6}A^2_3)\sqrt{\Ga_1},\label{m3I1}
\eea
where
\bea
\ga_1 &=&(1.4393\times 10^{-7}-1.03367\times 10^{-8}i)-(0.00196923-0.000141425i)A^2_3\crn
&+&(0.76764-0.0551299i)A^4_3,\label{ga1}\\
\ga'_1&=&(-0.00224904+0.000080656i)+(1.75343-0.0628823i)A^2_3,\label{gap1}\\
\Ga_1 &=&-7.94351\times 10^{12}+(6.19305\times 10^{15}-0.0625i)A^2_3\crn
&-&(7.05485\times 10^{15} +2.53004\times 10^{14}i)\sqrt{\ga_1}.\label{Ga1}
\eea

\end{document}